# PULSED LASER INTENSITY DEPENDENCE OF CRATER FORMATION AND LIGHT REFLECTION IN THE UDMA-TEGDMA COPOLYMER NANOCOMPOSITE, DOPED WITH RESONANT PLASMONIC GOLD NANORODS

*Ágnes Nagyné Szokol[1,2], Judit Kámán[1], Roman Holomb[1], Márk Aladi[1], Miklós Kedves[1], Béla Ráczkevi[1], Péter Rácz[1], Attila Bonyár[3], Alexandra Borók[3], Shereen Zangana[3], Melinda Szalóki[4], István Papp[1], Gábor Galbács, Tamás S. Biró[1,6,7], László P. Csernai[1,8,9], Norbert Kroó[1,10], Miklós Veres[1], NAPLIFE Collaboration*

[1]*HUN-REN Wigner Research Centre for Physics, 1121 Budapest, Hungary*
[2]*University of Pécs, Pécs, Hungary*
[3]*Department of Electronics Technology, Faculty of Electrical Engineering and Informatics, Budapest University of Technology and Economics, 1111 Budapest, Hungary*
[4]*Department of Biomaterials and Prosthetic Dentistry, Faculty of Dentistry, University of Debrecen, 4032 Debrecen, Hungary*
[5]*Department of Inorganic and Analytical Chemistry, University of Szeged, 6720 Szeged, Hungary*
[6]*Institute for Physics, Babeş Bolyai University, 3400 Cluj-Napoca, Romania*
[7]*Complexity Science Hub, 1080 Vienna, Austria*
[8]*Department of Physics and Technology, Bergen University, 5020 Bergen, Norway*
[9]*Csernai Consult, 5119 Bergen, Norway*
[10]*Hungarian Academy of Sciences*

**Abstract**

Plasmonic nanoparticles embedded into a solid matrix could play crucial role in laser-matter interactions. In this study, excess energy creation was observed during the single-shot irradiation of a polymer matrix containing plasmonic gold nanorods, resonant to the laser wavelength, with a high intensity femtosecond laser pulse. This effect was manifested in a 7-fold rise in the crater volume for a 1.7-fold increase of the laser intensity, and was absent in the pure polymer without the gold doping. It occurred at laser intensities $> 1.5 \times 10^{17}$ W/cm$^2$, being the vanishing threshold of plasma mirror formation, resulting in a more than 80% increase of the amount of laser light entering the target. This threshold was found to be critical for the plasmonic effect of gold nanoantennas tuned to the wavelength of the laser on the crater formation.

**Introduction**

There is an interesting tendency to concentrate research efforts to nuclear fusion, with the hope to satisfy our energy hungry world with environment friendly new resources [1,2,3]. The classical approach is to use techniques, which are based on equilibrium processes. However, these are struggling among others with instability problems, that can be eliminated by using non-equilibrium solutions. We intend to solve the problem by combining high-intensity laser fields and nanotechnology with the fusion process [4].

The possibility of laser-induced nuclear transitions has been proposed more than thirty years ago [5], and since then, such processes has been demonstrated experimentally for many different reactions [6,7,8], including fusion [9,10]. As it has been shown, laser intensities of $10^{19}$-$10^{21}$ W/cm$^2$, achievable with high-energy femtosecond lasers, could generate electric field gradients of the order $10^{14}$ V/cm, that are more than 100 times larger than the Coulomb field binding atomic electrons, creating favourable conditions for nuclear reactions.

However, the high laser intensities, required for the initiation of photonuclear reactions, can also be generated with lower energy laser pulses locally, by involving plasmonic nanostructures. Metallic nanoparticles of specific shape and size could have collective electron oscillations (plasmons i.e. plasmon polaritons) being resonantly coupled to the electromagnetic field of the laser. Under such circumstances these nanoparticles act as nanoantennas that amplify the laser field in their close vicinity by several orders of magnitude [11,12]. The plasmonic nanoparticles concentrate the light energy into hot spots on their surface, where the high field could accelerate the protons and explore the coherent, correlated plasmonic motion of the conduction electrons and the nearby heavier charged ions. The lifetime of these plasmons in the visible range of the light spectrum are in the few femtosecond range and there are powerful lasers (e.g.Ti:Sa) available in this range, with pulse length being optimal to excite the localized surface plasmons resonantly.

With the above-mentioned approach, the combination of ultrashort laser pulses and plasmonic nanoparticles allows to achieve local laser intensities of $10^{18}$-$10^{19}$ W/cm$^2$ with femtosecond lasers of moderate pulse energies in the range of tenth or hundreds of joules. The aim of the present work is to demonstrate this effect through the observation of excess energy generation during the resonant interaction of the plasmonic gold nanoparticles with high-intensity laser pulses. The peculiarities of crater formation in polymer targets (doped with plasmonic gold nanoparticles) upon their irradiation with these high-energy femtosecond laser pulses are studied, when the laser intensity on the target surface is varied by its longitudinal displacement from the focal plane.

**Methods**

*Sample preparation*

The materials and the sample preparation procedure are described in detail in ref. [11]. In short, two types of samples were prepared for these experiments: pure polymer (see below) target without gold nanoparticles (denoted as Au0), and that doped with gold nanorods (Au2). The size of the gold nanorods was selected so that they operate as plasmonic nanoresonators tuned to the 795 nm central wavelength of the femtosecond laser used for the irradiation experiments [13,14,15]. The polymer matrix was prepared from urethane dimethacrylate (UDMA, Sigma Aldrich Co., St. Louis, MO, USA) and triethylene glycol dimethacrylate (TEGDMA, Sigma Aldrich Co., St. Louis, MO, USA) in 3:1 weight ratio. 25 nm x 85 nm sized gold nanorods capped with dodecanethiol (Nanopartz Inc., USA, Product Number: B12-25-750-1DDT-TOL-50-0.25, Lot: L8969) were used for doping the UDMA-TEGDMA polymer with gold nanoparticles. The colloidal gold nanorods were added to the monomer mixture with ultrasound bathing for 5 minutes, resulting in a homogenized structure. The nominal concentration of the gold nanorods was $2.8 \times 10^{12}$ particles/mL, corresponding to 0.182 m/m%.

A droplet of the above-described mixture was placed between two glass slides, together with a plastic frame template (3D printed on one of the glass slides) ensuring the uniform gap. The formed discs were photopolymerized for 3 minutes from 5 cm distance through the glass slide by a standard dental curing lamp emitting blue light. The samples were prepared in ~240 µm thickness with ~1 cm diameter and were stored in a sample holder till the irradiation experiments.

*Laser irradiation experiments*

The single-shot irradiations of the UDMA-TEGDMA polymer targets were performed with a Coherent Hidra Ti:Sapphire chirped-pulse two-stage amplifier-laser system having a central wavelength of 795 nm. The laser pulse length was 42 fs. The diameter of the laser beam was 4 cm and it was focused onto the target with a lens of 400 mm focal distance. The angle of the laser beam and the target surface was 45°.

The samples were fixed to metal frame sample holders and placed into a vacuum chamber kept under $10^{-6}$ Pa pressure. The sample holder's position was changed with a motorized stage: it was moved after each shot laterally relative to the laser beam, in order to make each irradiation in an untreated region of the surface; and also longitudinally along the focused laser beam to change the laser intensity on the target surface. The irradiations were carried out with 27.7 mJ and 25 mJ pulse energies for the pure and doped with resonant gold nanorods UDMA-TEGDMA target, respectively.

*Reflection measurements*

The amount of the laser light reflected from the sample surface was detected in situ during the single-shot irradiations. A photodetector was arranged outside the vacuum chamber, behind one of its optical windows in 90° geometry so that it captured the laser beam reflected from the target surface.

*White light interferometry measurements*

The craters morphology and geometric parameters were investigated by a Zygo NewView$^{TH}$ 7100 white light interferometer. The instrument contains a single white light LED as illuminator with 580 nm central wavelength and 140 nm spectral widths. This configuration insures 3 μm coherence length for the 0.1 nm vertical resolution. During the measurements the scanning length was fitted to the studied surface – mostly between 20 μm and 65 μm.

The measurements were performed with a 50x magnification Mirau objective having an optical resolution of 0.52 μm. As the crater area was mostly over the objective's field of view (0.19 mm x 0.14 mm) it was a practical solution to use stitching, which insures the same resolution on a larger area. The stitching was carried out with 25-40% overlapping. After the measurements digital low pass filtering was applied to the experimental data. Then a reference plane was selected in the plane of the sample surface, and the volume of the crater was calculated from the surface map data [16]. The roughness parameters were also determined from the measured data.

## Results and Discussion

During the irradiation of a solid surface with a femtosecond high-energy laser pulse, crater formation occurs if the energy of the laser pulse transferred to the target material is above the ablation threshold. This phenomenon is affected by many factors, including the material of the solid, the laser wavelength, the pulse energy and intensity. The latter are interconnected, since the intensity on the target surface is determined by the ratio of the pulse energy and the laser spot area.

During the interaction of a high-intensity laser pulse ($>10^{13}$ W/cm$^2$) with a solid surface the ionization of the latter causes the formation of so-called plasma mirrors. These mirrors reflect the main part of a laser pulse specularly, preventing its absorption by the solid. While plasma mirrors are of great importance in handling high intensity laser pulses, they are a drawback when the aim of the laser irradiation is the processing, ablation or other interaction with the solid. In our experiments the aim was to transfer the highest possible amount of the energy of the laser pulse to the solid target, and to investigate the effect of plasmonic nanoparticles (having their plasmon resonance tuned to the laser wavelength) on this process.

As a first step, we studied the effect of plasma mirror formation by moving the target longitudinally across the focal plane of the laser and measuring the amount of reflected light from the target surface. During the displacement the laser spot size will change (it will be the smallest when the target surface is in the focal plane) resulting in different laser intensities on the sample from pulses of the same energy. The experiments were performed for both the doped and reference targets.

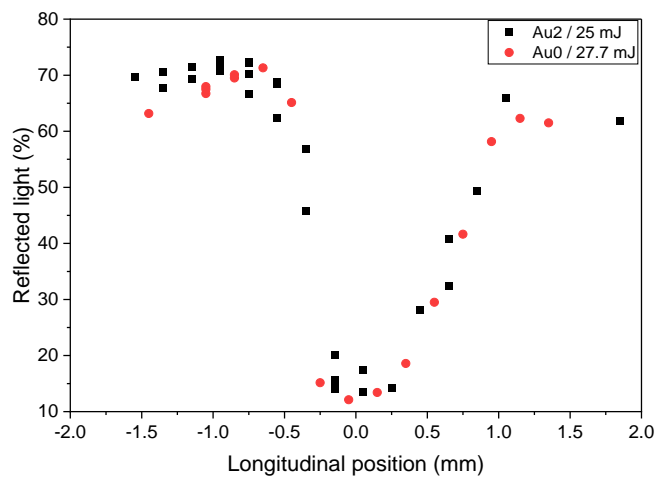

*Figure 1. Dependence of the reflected light on the longitudinal position of the target relative to the focused laser beam during the irradiation of the undoped (Au0) and Au doped (Au2) samples. Zero corresponds to the case when the surface was in the focus in both cases.*

As Figure 1 shows, in out-of-focus conditions (longitudinal displacement >1.0 mm from the focal point) a noticeable amount (>60%) of the incident light energy is reflected from the target. It can be seen that the reflection decreases slowly when moving the target further away from the focal point, and rapidly, when getting closer to it. When the target surface is in the focal point, the drop of the reflection is more than 80%. These observations are similar for the doped and undoped samples.

The above results can be interpreted in terms of the laser intensity (Fig. 2). For a given laser pulse energy and pulse length it is determined by the laser spot area, which, in turn, depends on the lateral displacement of the target from the focal plane. The laser intensities in our experiments are above $10^{16}$ W/cm$^2$, well in the plasma mirror formation region in all cases. As the intensity increases, up to $10^{17}$ W/cm$^2$, this effect is more pronounced, leading to increased reflection. However, above the intensity threshold of cca. 1.5 x $10^{17}$ W/cm$^2$ the plasma mirror vanishes, and the amount of laser energy transferred to sample rises significantly. So, achieving high-intensity irradiation conditions above the threshold ensures a more efficient energy transfer to the target. It can be seen that there are no significant differences between the data obtained on the undoped and gold-containing samples.

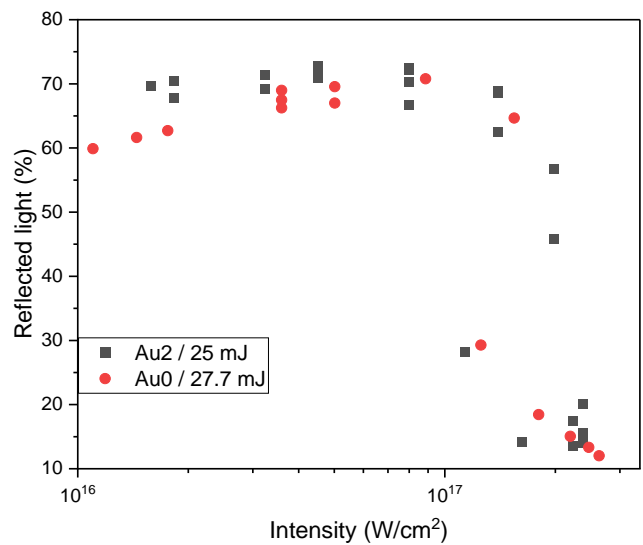

***Figure 2.*** *Change of the reflected light with the laser intensity for undoped (Au0) and gold nanorod containing (Au2) UDMA-TEGDMA targets.*

.

From Figures 1 and 2 it can be concluded that the doping of the target with plasmonic nanoparticles has no effect on the plasma mirror formation and the same decrease of the plasma mirror can be observed for both types of samples.

*Crater morphology*

Crater formation was observed during the single-shot irradiations along the whole studied longitudinal target position region, and in both the Au0 and Au2 samples. Figure 3 shows typical morphologies of the craters formed in doped and non-doped samples. It can be seen that the size and the morphology of the formed craters is influenced by the presence of gold nanorods. While the craters in the undoped samples have smooth crater walls, the surfaces of the gold-doped ones are rough. They feature small irregularities of a few microns size, that are especially characteristic for high laser intensities.

From the crater morphologies it can be concluded that the addition of plasmonic gold nanoparticles to the target influences the crater formation, resulting in the different crater wall morphology, and size. The volumes are also strikingly different, as displayed below in Figure 4.

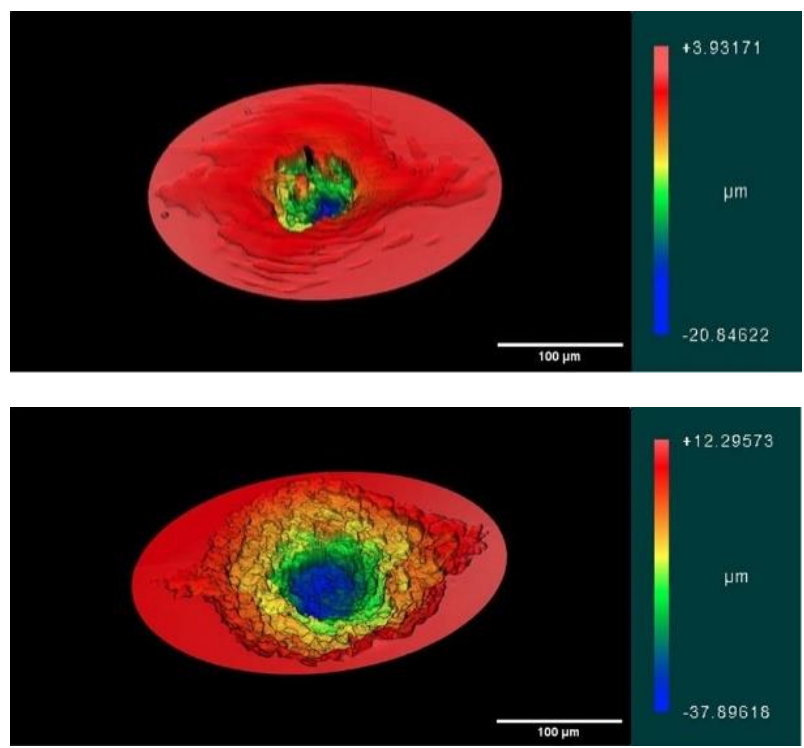

**Figure 3.** *White light interferometry images of the craters formed during the irradiation of the undoped (Au0, left) and the Au nanorod-doped (Au2, right) samples with about $2 \times 10^{17}$ W/cm$^2$ intensity laser pulses.*

*Crater volumes*

The volumes of the craters formed during the irradiation in the two types of samples with different laser intensities are shown in Fig. 4. As the Figure shows, up to around $10^{17}$ W/cm$^2$ the crater volume decreases with laser intensity following a linear (on the semilogarithmic scale) behaviour: it is $1.28 \times 10^5$ µm$^3$ for $1.1 \times 10^{16}$ W/cm$^2$ and $0.44 \times 10^5$ µm$^3$ for $8.8 \times 10^{16}$ W/cm$^2$ laser intensity.

However, the change of the crater volumes with laser intensity shows a completely different picture above cca. $1.5 \times 10^{17}$ W/cm$^2$. In the case of the undoped target the crater volume increases from $0.44 \times 10^5$ µm$^3$ to $1.13 \times 10^5$ µm$^3$, while the laser intensity changes from $0.89 \times 10^{17}$ W/cm$^2$ to $2.5 \times 10^{17}$ W/cm$^2$. So, a 2.8-fold increase in the laser intensity causes a 2.6-fold rise in the crater volume, meaning all the excess laser intensity entering the sample is converted into material removal.

The change of the crater volume is much more significant for the target doped with the plasmonic gold nanoparticles: from $0.77 \times 10^5$ µm$^3$ ($1.4 \times 10^{17}$ W/cm$^2$ laser intensity) it rises to $5.25 \times 10^5$ µm$^3$ ($2.4 \times 10^{17}$ W/cm$^2$ laser intensity). Here a 1.7-fold increase in the laser intensity leads to a 6.8-fold increase in the crater volume, indicating a non-linear relationship between the two quantities.

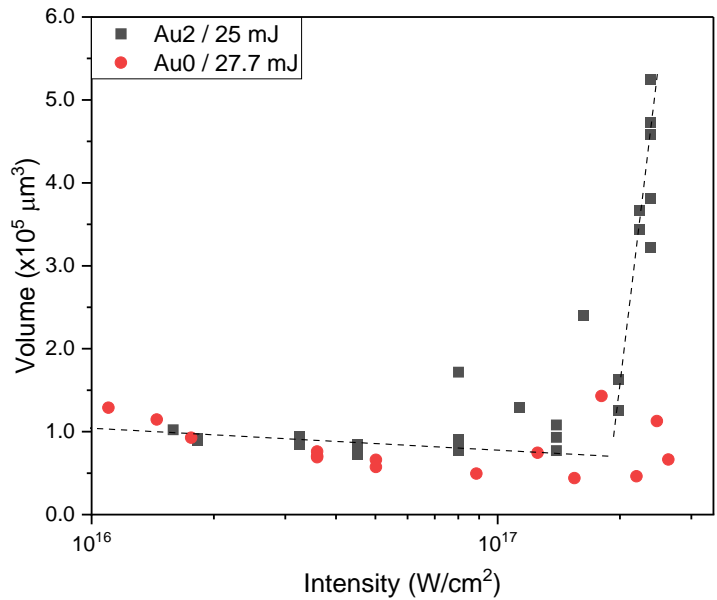

***Figure 4.*** *Change of the crater volume with the laser intensity for the laser irradiations of the undoped (Au0) and gold nanorod containing (Au2) targets.*

The rapid, almost 7-fold increase in the crater volume can be attributed only to the effect of the gold nanoparticles since those represent the only difference between the Au0 and Au2 targets. These nanorods, acting as nanoantennas, tuned to be resonant at the wavelength of the laser source, enhance the electromagnetic field of the latter in their vicinity and have some other advantages as described in one of our modelling papers [14]. This effect could increase the local intensity by several orders of magnitude, creating hotspots being able to initiate photoactivated nuclear processes.

The role of the gold nanoparticles is minimal below the $1.5 \times 10^{17}$ W/cm$^2$ threshold, but it is highly pronounced at higher laser intensities. Therefore, this threshold (where the plasma mirroring is vanished, so higher amount of the pulse enters the target (see Figure 5)) should be the limit above which the plasmonic effect and the amplification of gold nanoparticles could cause the formation of hotspots that could activate nuclear reactions.

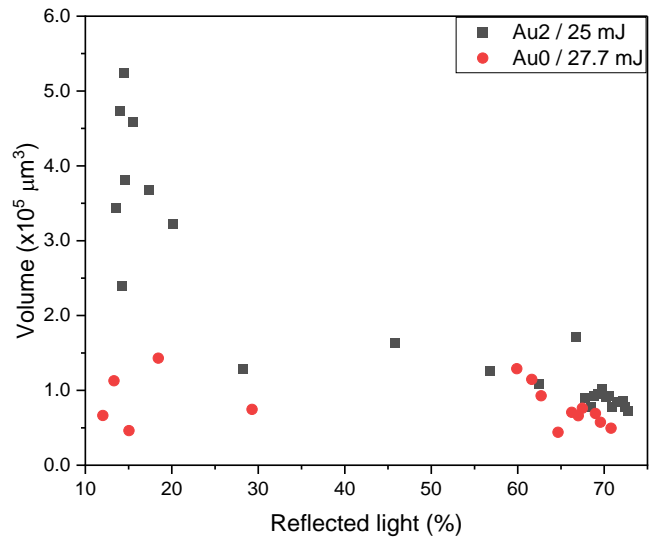

*Figure 5. Change of the crater volume with the amount of laser light reflected by the target (plasma mirror) for the laser irradiations of the undoped (Au0) and gold nanorod containing (Au2) targets.*

Proofs of the latter have been found during in-situ laser induced breakdown spectroscopy [17,18] experiments and the Raman spectroscopic studies of the crater walls [19], indication the possibility of hydrogen to deuterium reactions. In this process the nanorod antennas could catalyse energy production, possibly via nuclear transmutation reactions [20]. This is shown by the rapid increase of the crater volume even at constant laser pulse energy, but increasing intensity. Theoretical kinetic model simulations suggest that the embedded nanorod antennas exploit the collective resonant oscillation of massive number of electrons, which using the Laser Wake Field Acceleration (LWFA) mechanism accelerate the protons near the nanorod antennas to reach multi-MeV energies [21,22].

**Conclusions**

Excess energy creation was observed during the single-shot irradiation of a polymer matrix containing plasmonic gold nanorods, resonant to the laser wavelength, with high intensity femtosecond laser pulses. This effect, being absent in the pure polymer without the gold doping, occurred at laser intensities $> 1.5 \times 10^{17}$ W/cm$^2$, being above the threshold vanishing the plasma mirror formation. In this laser intensity region, the amount of laser light entering the sample is increased by more than 80%. This threshold is critical for the effect of the plasmonic gold nanoantennas tuned to the wavelength of the laser on the crater formation. Similar craters are formed in both undoped and doped targets and the crater volume decreases and correlates with the efficiency of the plasma mirror formation below $1.5 \times 10^{17}$ W/cm$^2$. In contrast, the volume increases with the laser intensity above the threshold, especially in the gold-doped target, where the 1.7-fold increase of the laser intensity is accompanied by an almost 7-fold rise in the crater volume.

This means, that additional energy has been created, which arises from processes involving the gold nanoparticles. Our suggestion is the occurrence of exotherm photonuclear processes, activated by the high-energy laser field enhanced by the plasmonic effects of the gold nanoparticles. The possibility of the hydrogen to deuterium reactions, as the explanation of this observation, is supported by our Raman spectroscopic and laser induced breakdown spectroscopy experiments. The nanoplasmonic activation mechanism is unique among the recent plethora of nuclear fusion ideas.

The hydrogen to deuterium transmutation has been detected only by optical spectroscopy methods. This finding has to be confirmed by nuclear detecting techniques as well as the determination of the details of the nuclear process. These studies are under way or in preparation.

**Acknowledgments**

This work is supported in part by the Hungarian Research Network, the Research Council of Norway, grant no. 255253, and the National Research, Development, and Innovation Office of Hungary, for projects: Nanoplasmonic Laser Fusion Research Laboratory under project Nr-s NKFIH-874-2/2020, NKFIH-468-3/2021, 2022-2.1.1-NL-2022-00002 and K 146733.

**References**

1 Mathew M.D., Nuclear energy: A pathway towards mitigation of global warming, Progress in Nuclear Energy 143 (2022) art. no. 104080. https://doi.org/10.1016/j.pnucene.2021.104080

2 Carayannis, E.G., Draper, J. & Bhaneja, B. Towards Fusion Energy in the Industry 5.0 and Society 5.0 Context: Call for a Global Commission for Urgent Action on Fusion Energy. J Knowl Econ 12 (2021) 1891–1904. https://doi.org/10.1007/s13132-020-00695-5

3 Barbarino, M. A brief history of nuclear fusion. Nat. Phys. 16, (2020) 890–893. https://doi.org/10.1038/s41567-020-0940-7

4 Biró, T.S.; Kroó, N.; Csernai, L.P.; Veres, M.; Aladi, M.; Papp, I.; Kedves, M.Á.; Kámán, J.; Nagyné Szokol, Á.; Holomb, R.; et al. With Nanoplasmonics towards Fusion. Universe 2023, 9, 233. https://doi.org/10.3390/universe9050233

5 K. Boyer, T. S. Luk, and C. K. Rhodes, Possibility of optically induced nuclear fission, Phys. Rev. Lett. 60 (1988) 557. https://doi.org/10.1103/PhysRevLett.60.557

6 K.W.D. Ledingham, Laser Induced Nuclear Physics and Applications, Nuclear Physics A 752 (2005) 633-644. https://doi.org/10.1016/j.nuclphysa.2005.02.132

7 Magill, J., Galy, J., Žagar, T. (2006). Laser Transmutation of Nuclear Materials. In: Schwoerer, H., Beleites, B., Magill, J. (eds) Lasers and Nuclei. Lecture Notes in Physics, vol 694. Springer, Berlin, Heidelberg . https://doi.org/10.1007/3-540-30272-7_9

8 Galy, J., Hamilton, D. & Normand, C. High-intensity lasers as radiation sources. Eur. Phys. J. Spec. Top. 175, 147–152 (2009). https://doi.org/10.1140/epjst/e2009-01133-4

9 V. S. Belyaev, A. P. Matafonov, V. I. Vinogradov, V. P. Krainov, V. S. Lisitsa, A. S. Roussetski, G. N. Ignatyev, and V. P. Andrianov, Observation of neutronless fusion reactions in picosecond laser plasmas, Phys. Rev. E 72, (2005) 026406. https://doi.org/10.1103/PhysRevE.72.026406

10 L. Giuffrida et al., High-current stream of energetic α particles from laser-driven proton-boron fusion, Phys. Rev. E 101 (2020) 013204. https://doi.org/10.1103/PhysRevE.101.013204

11 Bonyár, A.; Szalóki, M.; Borók, A.; Rigó, I.; Kámán, J.; Zangana, S.; Veres, M.; Rácz, P.; Aladi, M.; Kedves, M.Á.; et al. The Effect of Femtosecond Laser Irradiation and Plasmon Field on the Degree of Conversion of a UDMA-TEGDMA Copolymer Nanocomposite Doped with Gold Nanorods. Int. J. Mol. Sci. 2022, 23, 13575. https://doi.org/10.3390/ijms232113575

12 A. Bonyár, S. Zangana, T. Lednický, I. Rigó, I. Csarnovics, M. Veres, Application of gold nanoparticles–epoxy surface nanocomposites for controlling hotspot density on a large surface area for SERS applications, Nano-Structures & Nano-Objects 28 (2021) 100787. https://doi.org/10.1016/j.nanoso.2021.100787

13 Csete, M., Szenes, A., Tóth, E. et al. Comparative Study on the Uniform Energy Deposition Achievable via Optimized Plasmonic Nanoresonator Distributions. Plasmonics 17, 775–787 (2022). https://doi.org/10.1007/s11468-021-01571-x

14 Papp István, Bravina Larissa, Csete Mária, Kumari Archana, Mishustin Igor N., Motornenko Anton, Rácz Péter, Satarov Leonid M., Stöcker Horst, Strottman Daniel D., Szenes András, Vass Dávid, Szokol Ágnes Nagyné, Kámán Judit, Bonyár Attila, Biró Tamás S., Csernai László P., Kroó Norbert. Kinetic model of resonant nanoantennas in polymer for laser induced fusion. Frontiers in Physics 11. (2023). https://www.frontiersin.org/articles/10.3389/fphy.2023.1116023

15 L. P. Csernai, M. Csete, I. N. Mishustin, A. Motornenko, I. Papp, L. M. Satarov, H. Stöcker, N. Kroo (NAPLIFE Collaboration), Radiation dominated implosion with flat target, arXiv:1903.10896, Physics of Wave Phenomena, 28 (3), 187-199 (2020). https://doi.org/10.3103/S1541308X20030048

16 Ágnes Nagyné Szokol, for NAPLIFE Collaboration, Effect of the embedded plasmonic gold nanorods on the interaction of high intensity laser irradiation with UDMA polymer - the volume loss during crater formation, talk at 11th Int. Conf. on New Frontiers in Physics 2022, Kolymbari, Crete, Greece, 7th Sept. 2022.

17 N. Kroó, M. Aladi, M. Kedves, B. Ráczkevi, A. Kumari, P. Rácz, M. Veres, G. Galbács, L.P. Csernai, T.S. Biró, Monitoring of nanoplasmonics-assisted deuterium production in a polymer seeded with resonant Au nanorods using in situ femtosecond laser induced breakdown spectroscopy, arXiv:2312.16723 (2023) https://doi.org/10.48550/arXiv.2312.16723

18 LIBS Analysis of Pure, Deuterated and Au-doped UDMA: TEGDMA mixture: A Part of the Nanoplasmonic Laser Fusion Experiments, Archana Kumari, for the NAPLIFE Collaboration, 11th Int. Conf. on New Frontiers in Physics 2022, Kolymbari, Crete, Greece, 7th Sept. 2022.

19 I. Rigó, J. Kámán, Á. Nagyné Szokol, A. Bonyár, M. Szalóki, A. Borók, S. Zangana, P. Rácz, M. Aladi, M. Kedves, G. Galbács, L P. Csernai, T. S. Biró, N. Kroó, M. Veres, (NAPLIFE Collaboration), Raman spectroscopic characterization of crater walls formed upon single-shot high energy femtosecond laser irradiation of dimethacrylate polymer doped with plasmonic gold nanorods, arXiv: 2210.00619 [physics.plasm-ph], https://arxiv.org/abs/2210.00619

20 L. P. Csernai, I N. Mishustin, L M. Satarov, H Stöcker, L Bravina, M Csete, J Kámán, A Kumari, A Motornenko, I Papp, P Rácz, D D. Strottman, A Szenes, Á Szokol, D Vass, M Veres, T S. Biró, N Kroó (NAPLIFE Collaboration), Crater Formation and Deuterium Production in Laser Irradiation of Polymers with Implanted Nano-antennas, Phys. Rev. E, 108(2) 025205 (2023). https://doi.org/10.1103/PhysRevE.108.025205

21 István Papp, Larissa Bravina, Mária Csete, Archana Kumari, Igor N. Mishustin, Anton Motornenko, Péter Rácz, Leonid M. Satarov, Horst Stöcker, András Szenes, Dávid Vass, Tamás S. Biró, László P. Csernai, Norbert Kroó, Laser induced proton acceleration by resonant nano-rod antenna for fusion, arXiv: 2306.13445 [physics.plasm-ph], https://doi.org/10.48550/arXiv.2306.13445

22 L.P. Csernai, T. Csörgő, I. Papp, M. Csete, T.S. Biró and N. Kroó, New method to detect size, timespan and flow in nanoplasmonic fusion, arXiv: 2309.05156 [physics.plasm-ph], https://doi.org/10.48550/arXiv.2309.05156